\documentclass[twocolumn]{aastex631}\usepackage[T1]{fontenc}\usepackage{gensymb}
\usepackage{amsmath}
\usepackage{xspace}
\usepackage{color}
\usepackage{url}
\usepackage{hyperref}
\usepackage{graphicx}

\newcommand{\swift}{Swift J1727.8–1613}

\begin{document}

\title{A Comparison of the X-ray Polarimetric Properties of Stellar and Supermassive Black Holes}

\author[0000-0001-7163-7015]{M. Lynne Saade}
\affiliation{Science \& Technology Institute, Universities Space Research Association, 320 Sparkman Drive, Huntsville, AL 35805, USA}
\affiliation{NASA Marshall Space Flight Center, Huntsville, AL 35812, USA}

\author[0000-0002-3638-0637]{Philip Kaaret}
\affiliation{NASA Marshall Space Flight Center, Huntsville, AL 35812, USA}

\author[0000-0001-9200-4006]{Ioannis Liodakis}
\affiliation{NASA Marshall Space Flight Center, Huntsville, AL 35812, USA}
\affiliation{Institute of Astrophysics, Foundation for Research and Technology-Hellas, GR-70013 Heraklion, Greece}

\author[0000-0003-4420-2838]{Steven R. Ehlert}
\affiliation{NASA Marshall Space Flight Center, Huntsville, AL 35812, USA}

\begin{abstract}
 X-ray polarization provides a new way to probe accretion geometry in black hole systems. If the accretion geometry of black holes is similar regardless of mass, we should expect the same to be true of their polarization properties. We compare the polarimetric properties of all non-blazar black holes observed with IXPE. We find that their polarization properties are very similar, particularly in the hard state, where the corona dominates. This tentatively supports the idea that stellar and supermassive black holes share a common coronal geometry.

\end{abstract}

\section{Introduction}\label{sec:intro}

It is thought that stellar and supermassive black hole have similar accretion processes. Much evidence has been gathered that supports this hypothesis, such as the fundamental plane of black hole activity that relates X-ray luminosity, radio luminosity, and black hole mass \citep[e.g.][]{merloni2003,plotkin2012}, the broad-band radio luminosity \citep[e.g.][]{liodakis2017} of black hole jets, and direct comparisons of individual supermassive black holes with individual stellar black holes \citep[e.g.,][]{markoff2015,arcodia2020}. 

Accreting stellar and supermassive black holes show similar components to their spectra: a thermal component in the soft X-ray in stellar black holes and in the optical/UV in supermassive black holes, and a Comptonization component in the hard X-rays. The thermal component is attributed to the accretion disk, which is classically understood to be an optically thick, geometrically thin disk of gas like in \citet{shakura1973}. The Comptonized component is attributed to a corona, a hot plasma that inverse Compton scatters photons from the accretion disk to higher energies, creating the hard X-ray emission seen in black hole X-ray binaries (BHXBs) and active galactic nuclei (AGN) \citep{titarchuk1994}.

Obscuration is sometimes also observed in X-ray observations of black holes. This is attributed to a surrounding torus of dust and gas in AGN \citep[e.g.,][]{antonucci1993} or absorbing bulges in the outer edges of the accretion flow in dipping low-mass X-ray binaries \citep[e.g,][]{white1989,hirose1991} and obscured BHXBs \citep[e.g.][]{sutton2013,neilsen2020}. These features are thought to be quite vertically extended \citep[e.g.,][]{granato1997,buchner2014,neilsen2020}. When the line of sight to a source passes through these features, the direct X-ray emission from the inner accretion flow is obscured, and the observer sees scattered radiation. In unobscured sources, one can directly see the X-rays produced by the inner accretion flow.

It is well-understood that BHXBs go through different accretion states \citep{remillard2006,done2007,belloni2010}, the two primary states being the hard state and the soft state. In the hard state, the corona dominates the X-ray emission. The geometry of the corona is unknown. Among the geometries considered in the literature are a compact point source or sphere above the black hole \citep[lamppost, e.g.][]{wilkins2012}, a flat atmosphere of the accretion disk \citep[slab; e.g.,][]{haardt1993}, the base of a relativistic jet \citep[cone; e.g.,][]{henri1991,henri1997,markoff2005}, or associated with a geometrically thick, optically thin hot accretion flow \citep[wedge; e.g.,][]{tagliacozzo2023}. The latter might replace the geometrically thin optically thick disk in regions close to the black hole \citep{esin1997,done2007}, and is often hypothesized to be an advection-dominated accretion flow \citep[ADAF,][]{narayan1995}.

In the soft state the accretion disk dominates. The geometry of the inner accretion disk is thought to be different in the different states. In the soft state a geometrically thin optically thick accretion disk continues down to the innermost stable circular orbit (ISCO), while in the hard state the thin accretion disk is possibly truncated in its inner reaches, and possibly replaced with a geometrically thick, optically thin hot accretion flow close to the black hole \citep[e.g.,][]{done2007,yuan2014}. 

In between these states there exist intermediate states such as the hard intermediate and soft intermediate states \citep[e.g.,][]{belloni2010}, which are characterized by the presence of both substantial Comptonization and thermal spectral components. There is also the steep power law (SPL) state which is characterized by both thermal emission and a strong Comptonization component, but with a steeper powerlaw than the hard state \citep[e.g.,][]{dexter2014,hori2014}. 

State transitions were originally thought to be driven solely by changes in accretion rate, with the hard state primarily occurring at accretion rates 1\%-10\% of the Eddington limit, and the soft state primarily occurring at accretion rates 10\%-50\% of the Eddington limit \citep[e.g.][]{esin1997}. However, it was later found that different states can occur at the same luminosities \citep{remillard2006} and there is evidence that transitions between the states can occur at much lower luminosity ranges than previously thought \citep{maccarone2003}. It is generally true that when a BHXB is at a low luminosity, though, it is found in a hard state \citep{belloni2010}.

If black hole accretion functions similarly across the mass scale, then we should observe SMBHs to have accretion states as well. State transitions have been difficult to observe in AGN owing to the much longer viscous timescale, but we should be able to see different accretion states in different SMBHs, and there has indeed been evidence from population studies that AGN show accretion states like those of BHXBs \citep[e.g.,][]{kording2006,svoboda2017}. In recent years changing-look AGN have provided a potential look into state transitions in SMBHs through the observation of broad-band changes in the UV, X-ray, and optical spectrum \citep[e.g.,][]{noda2018}, and changes in the UV to X-ray spectral index \citep[e.g.][]{ruan2019}.

Polarization can provide an independent constraint on the geometry of the X-ray emitting region. X-ray polarization can therefore be used as a new and independent way to test how similar the accretion processes of stellar black holes and supermassive black holes truly are. For the first time, the Imaging X-ray Polarimetry Explorer \citep[IXPE,][]{weisskopf2022} allows X-ray polarization measurements of a wide range of stellar and supermassive black holes to be conducted.

The structure of this paper is as follows. In Section \ref{sec:indvs}, we review the polarization proprieties of all the non-blazar black hole sources observed with IXPE. In Section \ref{sec:plots}, we compare the polarization properties of the whole population of accreting black holes including stellar and supermassive objects and obscured and unobscured sources. In Section \ref{sec:discussion} we discuss the implications of our results, and in Section \ref{sec:conclusion} we state the conclusion of our research and what it means for future observations with IXPE.

\section{The X-ray Polarimetric Properties of the Individual Black Holes} \label{sec:indvs}
We review the X-ray polarization properties of the individual black holes in this section, as well as relevant quantities such as their inclination and Eddington ratios. For brevity when multiple papers on the same IXPE observation are available we focus on the papers produced by the IXPE collaboration. We use polarization degrees (PDs) and polarization angles (PAs) measured by the model-independent \texttt{PCUBE} \citep{baldini2022} algorithm unless otherwise specified. The system axis orientation is measured using the position angle of the radio jet or ionization cone of a source projected on the plane of the sky. PAs and position angles are measured counterclockwise from north. We assume a constant accretion efficiency of $\eta=0.1$, such that the bolometric luminosity is defined as $L_{bol}=\eta \dot{M} c^{2}$. With this assumption Eddington ratios in terms of luminosity and accretion rates are equivalent. We preferentially use Eddington ratios based on luminosity unless otherwise specified, as this is less model dependent than estimates of accretion rate based on spectral modeling. Inclinations represent the angle of the vector normal to the accretion disk plane with respect to the line of sight.

For clarity, we tabulate the polarization properties of the supermassive black holes in Table \ref{tab:SMBH}. We tabulate the polarization properties of the stellar black holes in Table \ref{tab:stellar}.

\subsection{Supermassive Black Holes}

\subsubsection{Circinus}
The Circinus galaxy is an active galaxy where the central accretion flow is obscured by a Compton-thick absorber with $N_{H}\sim 6\times10^{24}\;\mathrm{cm}^{-2}$ \citep{arevalo2014}. Its supermassive black hole has a mass $M_{BH}=(1.58^{+0.50}_{-0.55})\times10^{7}\;M_{\odot}$ \citep{koss2022b}. Its 
Eddington ratio is 0.03 \citep{phillipson2023}. Its inclination is uncertain; the inclination of the galaxy is 65\degr{} \citep{freeman1977} but the inclination of the AGN itself (i.e. the inclination of the accretion disk or jet as opposed to the disk of the galaxy) has been variously stated to be $i>70\degr$ based on the central 10pc CO(3-2) emission \citep{izumi2018}, $i>75\degr$ \citep{tristram2014} or $i>83$ \citep{isbell2022} based on the 1pc infrared disk, or close to 90\degr{} for the water maser disk \citep{greenhill2003,izumi2018,isbell2022}.

The IXPE observation of Circinus was analyzed by \citet{ursini2023}. The average 2-8 keV PD was found to be $28\pm7\%$, and the PA was $18\pm5\degr$. The PA is perpendicular to the radio jet of the AGN, which has a position angle of 295\degr \citep{elmouttie1998}. \citet{ursini2023} note the polarization properties are consistent with expectations for reflection off of the AGN torus. They calculate that the torus is thick, with a height to radius ratio $\sim1$ and a covering fraction of 0.7.

\subsubsection{NGC 1068}
NGC 1068 is an active galaxy classified as a Compton-thick, with $N_{H}\sim10^{25}\;\mathrm{cm}^{-2}$ \citep{bauer2015}. Its supermassive black hole has a mass $(3.16^{+0.64}_{-0.53})\times10^{6}\;M_{\odot}$ \citep{davis2023}. Its Eddington ratio is 0.03 \citep{koss2022a}. Various studies have measured its inclination to be $70\degr-90\degr$ \citep[e.g.,][]{honig2007,honig2008,lopez-rodriguez2018} .

Its IXPE observation was analyzed by \citet{marin2024}. The average 2-8 keV PD was $12.4\pm3.56\%$ and the PA was $101\pm8\degr$. The PA was roughly perpendicular to the jet position angle, which is $\sim24\degr$ \citep{wilson1983,mutie2024}. Overall \citet{marin2024} state that NGC 1068 has polarization properties similar to those of Circinus, with the high polarization presumably being attributed to reflection off of the AGN torus. They estimate the torus has a half-opening angle of 50-55\degr.

\subsubsection{NGC 4151}

NGC 4151 is an unobscured Seyfert galaxy. It has an SMBH mass $M_{BH}=(1.66^{+0.48}_{-0.34})\times10^{7}\;M_{\odot}$ \citep{bentz2022}. Its accretion rate is 0.02$ \dot{M}_{Edd}$ \citep{li2022}. The inclination of the AGN in NGC 4151 is very uncertain \citep{marin2016}. It has been estimated to be $0<i<33\degr$ when using the X-ray relativistic reflection component \citep[e.g.,][]{nandra1997,keck2015,beuchert2017,miller2018} or $\sim58\degr$ using broad-line reverberation studies \citep{bentz2022}. The IXPE observation of NGC 4151 was analyzed by \citet{gianolli2023}. They measured an average 2-8 keV PD of $4.9\pm1.1\%$ and a PA of $86\pm7\degr$. They note that the PD is too high to be explained by a corona with a lamppost or conelike geometry. These geometries are further disfavored by the PA being closely aligned with the extended radio emission of NGC 4151, which has a position angle of $\sim83\degr$ \citep{harrison1986,ulvestad1998}. The authors state slab and wedge-shaped coronal geometries are favored for the source.

\subsubsection{IC 4329A}

IC 4329A is an unobscured Seyfert galaxy \citep{veron-cetty2006}. It has an SMBH mass $M_{BH}=(6.8^{+1.2}_{-1.1})\times10^{7}\;M_{\odot}$ \citep{bentz2023}. Its Eddington ratio is 0.06 \citep{caglar2023}. IXPE observed this AGN on 2023 Jan 5-15, and this observation was analyzed by \citet{ingram2023_IC}. They detected an average 2-8 keV PD of $3.3\pm1.1\%$, though this was technically only a $2.97\sigma$ detection. The measured PA was $78\pm10\degr$. The PA is possibly aligned with the radio jet, as the radio image of the source shows a small-scale position angle of 78\degr \citep[though at larger scales the position angle is closer to 90\degr;][]{unger1987}. \citet{ingram2023_IC} note that an asymmetric, outflowing corona extended in the disk plane is favored. Reflection modeling constrains the inclination of the inner disk to be $<39\degr$ at 99\% confidence.

\subsubsection{MCG 05-23-16}

MCG 05-23-16 is a Seyfert galaxy classified as unobscured \citep{veron1980}. It has an SMBH mass of $M_{BH}=(4.47^{+0.45}_{-0.39})\times10^{7}\;M_{\odot}$ \citep{koss2022b}. Its Eddington ratio is 0.02 \citep{garcia2022}. It was observed by IXPE twice, the first observation analyzed by \citet{marinucci2022}, and the second observation analyzed by \citet{tagliacozzo2023}. It has an inclination $<60\degr$\citep{zoghbi2017}.

In the first observation, only a 99\% confidence upper limit on the 2-8 keV PD was measured, a limit of $<4.7\%$. \citet{marinucci2022} stated this was consistent with a spherical lamppost or conical coronal geometry, while if the geometry of the corona was a slab above and below the disk, the inclination had to be $<50\%$. 

In the second observation, another upper limit was obtained, this time $<3.2\%$. The 2-8 keV PA was $53\pm13\degr$, which \citet{tagliacozzo2023} noted was potentially aligned with the narrow-line-region (NLR) axis of this source, which is approximately 40\degr \citep{ferruit2000}. This potentially hints that the corona is extended in the disk plane. \citet{tagliacozzo2023} cite work in preparation by Serafinelli et al. that uses the reflection component to constrain the inclination to be $30\degr<i<50\degr$. If that is correct, a cone-shaped geometry for the corona can be ruled out.

\begin{deluxetable*}{lccccccc}
\tablecaption{Polarization information of supermassive black holes in the IXPE sample. \label{tab:SMBH}}
\tablewidth{2pt}
\tablehead{
\colhead{Name} & 
\colhead{State} &
\colhead{Mass $(M_{\odot})$} &
\colhead{$L_{bol}/L_{Edd}$} &
\colhead{PD (\%)}&
\colhead{PA (\degr)}&
\colhead{System axis angle (\degr)}&
\colhead{Inclination (\degr)}}
\startdata
Circinus & obscured & $(1.58^{+0.50}_{-0.55})\times10^{7}$ & 0.03 & $28\pm7$ & $18\pm5$ & 295 & 70-90 \\
NGC 1068 & obscured & $(3.16^{+0.64}_{-0.53})\times10^{6}$ & 0.03 & $12.4\pm3.56$ & $101\pm8$ & 24 & 70-90 \\
NGC 4151 & hard & $(1.66^{+0.48}_{-0.34})\times10^{7}$ & 0.02 & $4.9\pm1.1$ & $86\pm7$ & 83 & 0-58\\
IC 4329A\tablenotemark{a} & hard & $(6.8^{+1.2}_{-1.1})\times10^{7}$ & 0.06 & $3.1\pm1.1$ & $78\pm10$ & 78 & $<39$\\
MCG 05-23-16 & hard & $(4.47^{+0.47}_{-0.39})\times10^{7}$ & 0.02 & $<3.2$ & $53\pm13$ & 40 & 0-60 
\enddata
\tablenotetext{a}{Only $2.97\sigma$ detection}
\tablecomments{PAs and System Axis Angles are measured counterclockwise from north. Classification of some SMBH sources as "hard" is based solely on Eddington ratio.}
\end{deluxetable*}

\subsection{Stellar Mass Black Holes}
\subsubsection{Cygnus X-1}
Cygnus X-1 is a stellar black hole X-ray binary (BHXB) with mass $M_{BH}=21.2\pm2.2\;M_{\odot}$  and an orbital inclination of $27.5\pm0.8\degr$ \citep{miller-jones2021}. Fitting of its spectra implies a higher inclination for the axis of the system, $40\degr-70\degr$ \citep{tomsick2014,walton2016}. It was observed by IXPE twice in the hard state and five times in the SPL state. A detailed analysis of the first hard state IXPE observation was performed by \citet{krawczynski2022}. The time-averaged PD in the 2-8 keV band was $4.01\pm0.2\%$ while the PA was measured to be $339.3\pm1.4\degr$. The polarization degree increased with energy, implying the Comptonizing corona is the primary source of the polarized signal. The position angle of the radio jet on the sky was $338\degr$, making the PA aligned with the radio jet. \citet{krawczynski2022} argued this implied the corona in Cyg X-1 does not have the lamppost geometry, favoring instead a slablike corona sandwiching the accretion disk or a hot accretion flow in the inner parts of the disk. A conelike geometry of the corona along the jet axis is also disfavored by the data, as it predicts a PD lower than observed. The relatively high polarization degree could be further evidence that the inclination of the X-ray emitting region is greater than the $27\degr$ inclination of the binary orbit \citep{krawczynski2022,zdiarski2023}. Alternatively it could imply a mildly relativistic outflow from the corona \citep{poutanen2023} or jet \citep{dexter2024}. The Eddington ratio of Cyg X-1 in this state was 0.01.

\citet{steiner2024} report an analyis of 5 IXPE observations taken when Cygnus X-1 was in a softer state identified as the SPL state. The time-averaged PD in the 2-8 degree band is $1.9\pm0.13\%$ in the SPL state. The PD is lower in the SPL state than in the hard state. The PA is $334.3\pm1.8\degr$, which is aligned to the system axis, as in the hard state. This is naively what we expect, as the SPL state of Cygnus X-1 is dominated still by the corona, and not by the disk as in the canonical soft state of X-ray transients \citep{grinberg2013}. \citet{steiner2024} attribute the polarization in the SPL state to returning radiation, however, so the similar polarization properties might be coincidental. The Eddington ratio of Cygnus X-1 in this state was 0.006, based on the accretion rate.

\subsubsection{Cygnus X-3}
Cygnus X-3 is an X-ray binary with an orbital inclination of $29.5\pm1.2\degr$ \citep{antokhin2022} and a jet inclination of $<14 \degr$ \citep{mioduszewski2001}. The compact object is low in mass $(M=2.4^{+2.1}_{-1.1}\;M_{\odot})$ , making it possible it is a neutron star, but its radio, X-ray, and infrared properties suggest it is more likely to be a black hole \citep{zdziarski2013}. Despite its low inclination it is also considered an obscured source, with evidence for this including dips in the X-ray light curve \citep{1982ApJ...257..318W}, a low energy for the cutoff in its hard state power law, and a lack of power at high frequencies in its power spectrum \citep{Zdziarski2010}.

Cygnus X-3 was observed by IXPE once in a harder state, and once in a softer state. These observations were analyzed by \citet{veledina2023_cyg}. In the harder state observations they found a high 2-8 keV average PD $20.6\pm0.3\%$ and a PA of $90.1\pm0.4\degr$. The soft state observations had average 2-8 keV PD $10.4\pm0.3$\% and a PA of $92.6\pm0.6\degr$. The position angle of the jet is $4\pm2\degr$ \citep{marti2000}, making the PA orthogonal to the jet in both the hard and soft states. The observed polarization properties are similar to those of the Circinus AGN \citep{ursini2023}, which is obscured by a thick torus of gas with an opening angle smaller than the inclination of the line of sight to the AGN. Only X-rays reflecting off of the torus are observed.  \citet{veledina2023_cyg} argue that Cyg X-3 must have a similar geometry, such that we are only seeing X-rays reflected off of narrow funnels of gas that surround the source. This further implies the luminosity of the source is much higher than first impressions suggest given that the primary radiation is heavily absorbed, making Cyg X-3 an ultraluminous X-ray source (ULX).

\subsubsection{Swift J1727.8–1613}
Swift J1727.8–1613 is a transient low-mass BHXB discovered on 2023 Aug 24 when it first went into outburst \citep{kennea2023,negoro2023}. Its mass is $M_{BH}=10\pm2\;M_{\odot}$ and its inclination is $38\pm3\degr$ according to X-ray spectral fitting, though it should be noted the mass and inclination are degenerate in this method \citep{svoboda2024_swift}. It was first observed by IXPE in the hard state \citep{veledina2023_swift} and successive observations covered the transition to the hard intermediate state \citep{ingram2023_swift}. Further observations were done when the source was in the soft state in \citep{svoboda2024_swift}, and when it entered the hard state again, though at a much lower luminosity \citep[low hard state or LHS;][]{podgorny2024}

\citet{veledina2023_swift} reported a time-averaged PD in the 2-8 keV band of $4.1\pm0.2\%$ using spectropolarimetric fitting of the IXPE data, and a time-averaged PA in the same band of $2.2\pm1.3\degr$. The PD increased with energy, as is consistent with the polarization arising from Comptonization. The jet in the hard state was long and continuous, with a position angle $359.4\pm0.07\degr$ \citep{wood2024}. The X-ray PA is thus closely aligned to the system axis. \citet{veledina2023_swift} argue that this indicates the X-ray emission region is extended in the accretion disk plane and orthogonal with the jet. Together with the increasing PD with energy, this disfavors spherical or lamppost geometries of the corona, similar to Cyg X-1. 

\citet{ingram2023_swift} present a further analysis including not only the first IXPE observation of \swift{} but the observations covering the transition to the hard intermediate state. The time-averaged 2-8 keV PD and PA for the source was measured using spectropolarimetric fits of the IXPE and NICER data. A fit where the PD increased with energy is preferred. The PD steadily declines from $\sim4\%$ in the first couple of observations to $\sim3\%$ in the last couple of observations, decreasing as the source becomes softer. In contrast the PA remains roughly constant throughout the state transition, implying it remains aligned with the system axis. This leads to similar conclusions about the coronal geometry to \citet{veledina2023_swift}. 

The two IXPE observations of the source in the soft state were analyzed by \citet{svoboda2024_swift}. The time-averaged 2-8 keV PD was only an upper limit, $<1.2\%$ at 99\% confidence. The reported PA was $8\pm21\degr$. However, since polarization is not confidently detected, this result should be treated with caution. The authors argue the drop in the PD as the source evolves along the q-diagram favors a scenario in which changes in the X-ray polarization signal are driven by changes in the innermost accretion flow. The observed very low PD is consistent with theoretical predictions for pure thermal emission from a geometrically thin and optically thick disk, as is expected to make up the inner accretion flow for the soft state of a BHXB.

\citet{podgorny2024} present an analysis of \swift{} in the low hard state. The time-averaged 2-8 keV PD was $3.3\pm0.4\%$, and the PA was $3\degr\pm4\degr$. These match the polarization properties of the source in the initial hard state observation, when the source was at a higher luminosity and rising in luminosity. They note the polarization properties are also very similar to NGC 4151. Their simulations indicate that if the corona had a slab geometry, it could replicate the polarization properties of the hard state, even at the two very different luminosities observed for this source. However if the slab was located on the optically thick, geometrically thin disk as opposed to a hot inner accretion flow, they would expect a strong energy dependence of the PD, and a rotation of the PA with energy, neither of which is observed.

\subsubsection{LMC X-1}
LMC X-1 is a BHXB in the Large Magellanic Cloud with a black hole $10.91\pm1.41M_{\odot}$ in mass and an orbital inclination of $36.38\pm1.92\degr$ \citep{orosz2009}. It was observed by IXPE in the soft state, and this observation was analyzed by \citet{podgorny2023}. The time-averaged 2-8 keV PD was only an upper limit, 1.1\% at 99 percent confidence. The reported PA was $51.6\pm11.8\degr$. However since polarization was not confidently detected, the result should be interpreted with caution. The reported PA roughly aligns with the source's ionization cone \citep[position angle 50\degr][]{cooke2007,cooke2008}. The observed very low PD is consistent with theoretical predictions for BHXBs in the soft state \citep{podgorny2023}. The alignment of the PA with the ionization cone suggests that a slab corona is present in the system, similar to Cyg X-1 and \swift{}, though the lack of a robust polarization detection means these conclusions are tentative. Its Eddington ratio during the observation was 0.11, based on the accretion rate derived from a model fit to the spectrum and assuming a black hole mass of 10.9$M_{\odot}$.

\subsubsection{LMC X-3}
LMC X-3 is a BHXB in the Large Magellanic Cloud with a black hole about $6.98\pm0.56M_{\odot}$ in mass and an orbital inclination of 69.2\degr \citep{orosz2014}. IXPE observed LMC X-3 on while it was in the soft state \citep{svoboda2024_LMC}. \citet{svoboda2024_LMC} report an average 2-8 keV PD of $3.1\pm0.4\%$ and a PA of $315\pm4\degr$. The PD is constant below 5 keV, and increases above 5 keV with a confidence level of only 1$\sigma$. The authors state that the level of the PD is consistent with expectations for a thermal accretion disk around a slowly spinning black hole with high inclination. No radio jet has ever been detected from this source \citep{fender1998,gallo2003,lang2007} and no ionization cones have been detected either \citep{hutchings2003}. It is therefore not possible to compare the PA to the orientation of the system axis. During the observations, its Eddington ratios were in the 0.4-0.45 range.

\subsubsection{4U 1957+115}
4U 1957+115 is a low mass x-ray binary (LMXB) that is consistently observed in a soft state \citep[e.g.][]{yaqoob1993,ricci1995,nowak1999,nowak2008,nowak2012,maitra2014,sharma2021,barillier2023}.  Its mass has not been measured directly but it is most likely a black hole based on the X-ray properties \citep{maccarone2020}, and \citet{gomez2015} estimate the mass is <6.2 $M_{\odot}$ with 90\% confidence. Its inclination is uncertain. No eclipses or X-ray light curve orbital modulations or highly ionized winds have been observed, implying the inclination is $<65-75\degr$ \citep{wijnands2002,ponti2012,parra2024}. However, X-ray spectral fitting predicts $i\sim78\degr$ \citep{maitra2014} and some optical modulation studies favor low inclinations $\sim13\degr$ \citep{gomez2015}. 

The IXPE observation of 4U 1957+115 was analyzed by \citet{marra2023}. The average PD in the 2-8 keV band was $1.9\pm0.4\%$ and the PA was $355.8\pm5.2\degr$. There was little to no time variability of the polarization properties throughout the observation. The 2-3, 3-4.3, and 4.3-6 keV bands show an increase in PD with energy, while the polarization is only an upper limit in the 6-8 keV band. The authors argue the increasing PD with energy can be explained by a strong contribution from returning radiation. Since this source stays in the soft state a jet has never been observed, and there is no other source of information about the system axis (such as ionization cones). Overall \citet{marra2023} concluded the polarization properties of 4U 1957+115 are consistent with theoretical expectations for an optically thick and geometrically thin accretion disk around the black hole. Its Eddington ratio during this observation was 0.02, using the accretion rate measured from a model fit to the spectrum and an assumed black hole mass of 4.6 $M_{\odot}$.

\subsubsection{4U 1630-47}
4U 1630-47 is a transient LMXB. It has an inclination of $60\degr-70\degr$ due to the presence of X-ray dips in its light curve, but a lack of eclipses \citep{1998ApJ...494..753K,tomsick1998}. The mass is poorly constrained, but its spectral properties suggest the compact object is a black hole \citep{parmar1986}, and \citep{ratheesh2024} estimate a mass of $18^{+0.7}_{-1.2}\;M_{\odot}$ using spectral fitting to NICER and NuSTAR data. It has been observed by IXPE in a high-luminosity soft state (the high soft state; HSS) \citep{ratheesh2024} and SPL states \citep{rodriguez2023}.

\citet{ratheesh2024} report a 2-8 keV average PD of $8.32\pm0.17\%$ and an average PA of $17.8\pm0.6\degr$ from the source in the HSS. The PD increases with energy (up to 10\% at 8 keV) while the PA remains constant with energy. The authors argue that standard models of thin accretion disks cannot explain such a high polarization, as general relativistic effects would reduce the polarization coming out of the disk to lower values than observed. They argued polarimetric properties can be explained if the thin disk has an outflowing atmosphere, with a high optical depth of $\tau \sim 7$ and a velocity $\sim0.5c$, though the authors note that there might be other accretion flow geometries they have not considered that could give rise to the polarization. A distant wind is disfavored as a source of the polarization. There is no jet observed for this source, so the measured PA cannot be compared to the system axis. The Eddington ratio during this observation was around 0.5.

\citet{rodriguez2023} report a 2-8 keV average PD of $6.8\pm0.2\%$ and an average PA of $21.3\pm0.9\degr$ for 4U 1630-47 in the SPL state. They state the small change in the PA from the soft state to the SPL state is surprising given the very different energy spectra, as is the relatively high PD continuing into the steep power law state. This implies the geometry in the SPL is similar to that in the HSS. As with the high soft state observations, standard models of thin accretion disks cannot explain the high polarization observed. The Eddington ratio during this observation was around 0.12, based on a mass accretion rate estimated from a model fit to the spectrum and a black hole mass of 18 $M_{\odot}$.

\begin{deluxetable*}{lccccccc}
\tablecaption{Polarization information of stellar black holes in the IXPE sample. \label{tab:stellar}}
\tablewidth{2pt}
\tablehead{
\colhead{Name} & 
\colhead{State} &
\colhead{Mass $(M_{\odot})$} &
\colhead{$L_{bol}/L_{Edd}$} &
\colhead{PD (\%)} &
\colhead{PA (\degr)} &
\colhead{System axis angle (\degr)} &
\colhead{Inclination (\degr)}}
\startdata
Cyg X-1 & hard & $21.2\pm2.2$ & 0.01 & $4.01\pm0.2$ & $339.3\pm1.4$ & 338 & 27.5-70 \\
{} & SPL & {} & 0.006 & $1.9\pm0.13$ & $334.3\pm1.8$ & {} & {}\\
Cyg X-3 & obscured (harder) & $2.4^{+2.1}_{-1.1}$ & unknown & $20.6\pm0.3$ & $90.1\pm0.4$ & 4 & $<14$\\
{} & obscured (softer) & {} & {} & $10.4\pm0.3$ & $92.6\pm0.6$ & {} & {}\\
Swift J1727.8–1613 & hard & $10\pm2$ & not given & $4.1\pm0.2$ & $2.2\pm1.3$ & 359 & $38.3\pm3$\\
{} & soft & {} & not given & $<1.2$ & $8\pm21$ & {} & {}\\
{} & low hard & {} & not given & $3.3\pm0.4$ & $3\pm4$ & {} & {}\\
LMC X-1 & soft & $10.91\pm1.41$ & 0.11 & $<1.1$ & $51.6\pm11.8$ & 50 & $36.38\pm1.92$\\
LMC X-3 & soft & $6.98\pm0.56$ & 0.4-0.45 & $3.1\pm0.4$ & $315\pm4$ & unknown & 69.2\\
4U 1957+115 & soft & <6.2 & 0.02\tablenotemark{a} & $1.9\pm0.4$ & $355.8\pm5.2$ & unknown & 13-78\\
4U 1630-47 & soft & $18^{+0.7}_{-1.2}$ & 0.5\tablenotemark{b} & $8.32\pm0.17$ & $17.8\pm0.6$ & unknown & 60-70\\
{} & SPL & {} & 0.12\tablenotemark{b} & $6.8\pm0.2$ & $21.3\pm0.9$ & {} & {}\\
\enddata 

\tablenotetext{a}{Assuming a mass $4.6\;M_{\odot}$}
\tablenotetext{b}{Assuming a mass $18\;M_{\odot}$}
\tablecomments{PAs and System Axis Angles are measured counterclockwise from north. Eddington ratios assume constant accretion efficiency with luminosity.}
\end{deluxetable*}

\section{Black Hole Polarization Comparison}\label{sec:plots}

To compare the polarization properties of the stellar and supermassive black holes, we plot their polarization properties along side each other in a series of plots. This allows us to investigate whether the polarization properties, and therefore the accretion geometries, are similar.

In the following plots, orange denotes a soft state source, blue denotes a hard state source (or a source predicted by theory to be hard state, in the case of AGN), and black denotes a source that is known to be obscured. Magenta denotes a source in the steep power law state. All of these state classifications are done independently of the source's polarization properties (e.g. obscured sources are categorized as obscured based on their spectra, not on the polarization degree). Stellar mass black holes are plotted as circles, supermassive black holes are plotted as squares. The low hard state of \swift{} is plotted separately from its earlier hard state observation, as the latter occurred at a higher luminosity.

\begin{figure*}
\centering
\includegraphics[width=\textwidth]{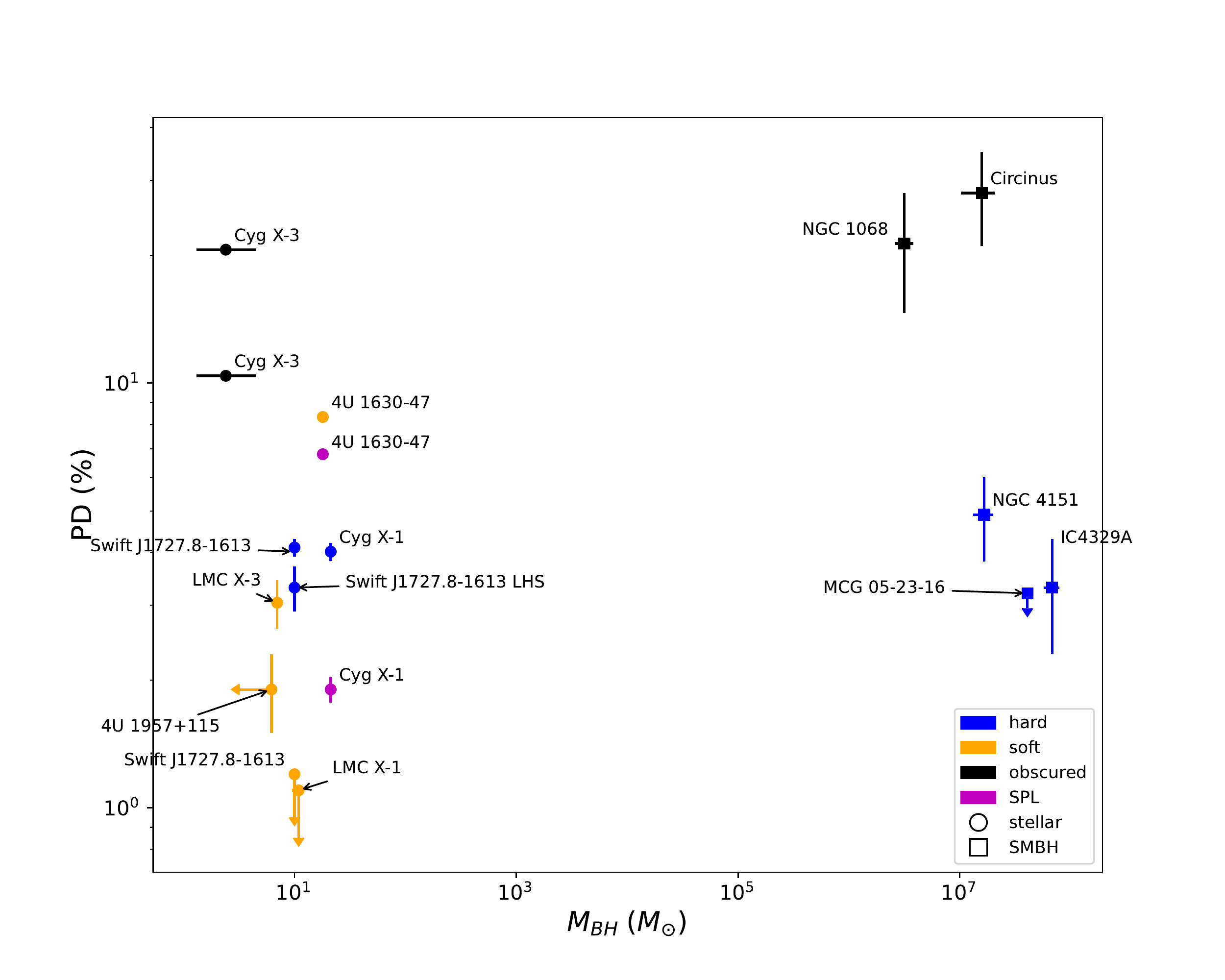}
\caption{Plot of the logarithm of the PD vs the logarithm of the black hole mass $(M_{BH})$. Orange denotes a soft state observation, blue denotes a hard state observation, magenta denotes an SPL observation, and black denotes a source that is obscured. Stellar mass black holes are plotted as circles, supermassive black holes are plotted as squares. Note that MCG 05-23-16, the soft state of \swift{}, and LMC X-1 have upper limits on the PD, while 4U 1957+115 has an upper limit on the mass. There is no clear trend in the PD with respect to mass, as expected if the accretion geometry of stellar black holes and supermassive black holes is similar.}
\label{fig:PD_vs_mass}
\end{figure*}

Figure \ref{fig:PD_vs_mass} plots the PD of each source in the 2-8 keV range versus its black hole mass, with the exception of NGC 1068, for which PD is plotted in the 2-6 keV range to exclude the unpolarized Fe K-alpha line \citep[see][]{marin2024}. Both axes are in logarithmic scale. MCG 05-23-16 and LMC X-1 have only upper limits on the PD, while 4U 1957+115 only has an upper limit on the mass. There is no clear trend in PD with respect to mass. This is what is expected if the accretion geometries of black holes are similar across the mass scale.

\begin{figure*}
\centering
\includegraphics[width=\textwidth]{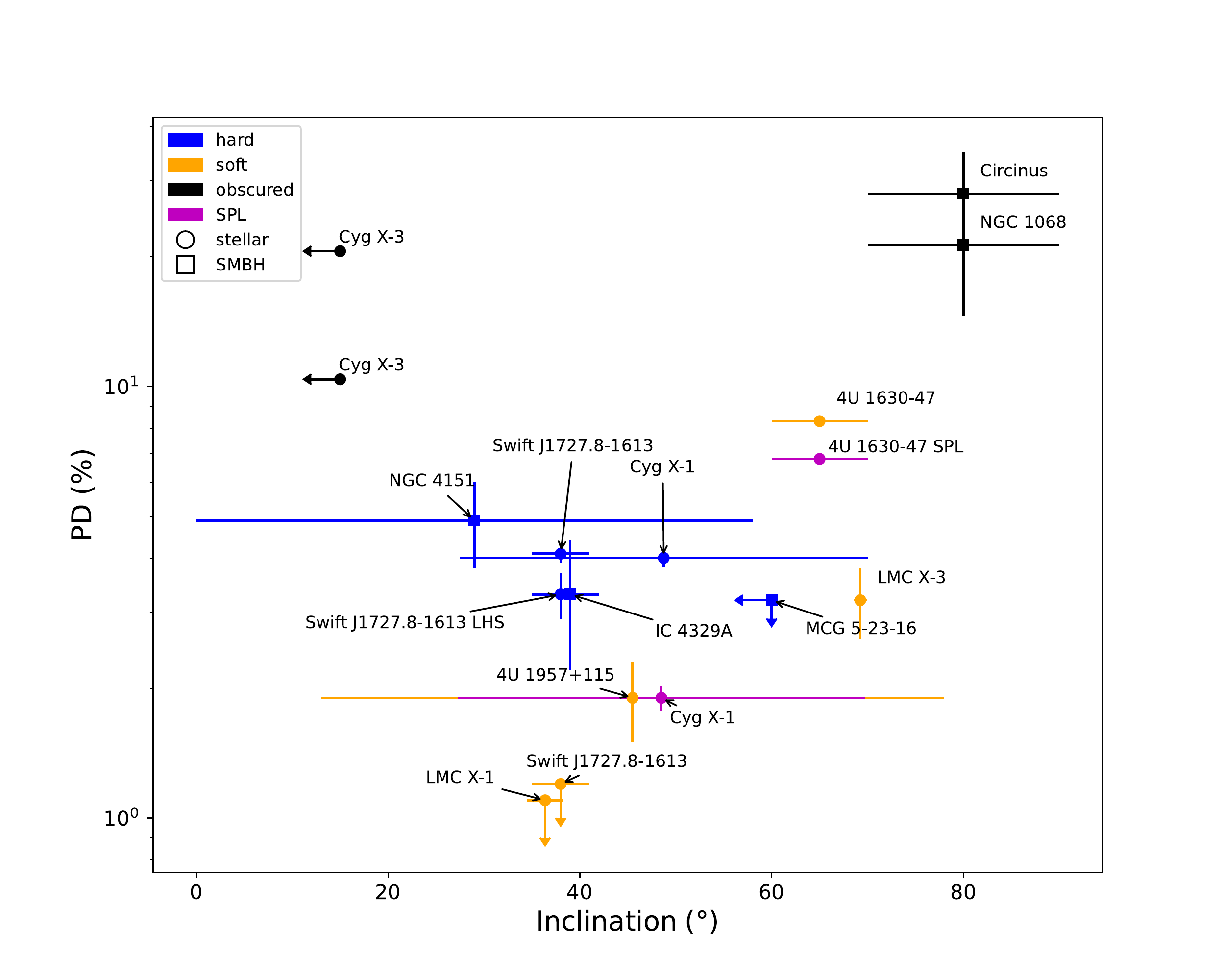}
\caption{Plot of the PD of the IXPE black hole sources vs their inclination. Orange denotes a soft state observation, blue denotes a hard state observation, magenta denotes an SPL observation and black denotes an obscured source. Stellar mass black holes are plotted as circles, supermassive black holes are plotted as squares. Note that IC 4329A and MCG 5-23-16 have upper limits on the inclination.}
\label{fig:PD_vs_inclination}
\end{figure*}

Figure \ref{fig:PD_vs_inclination} plots the PD of each source versus its inclination. The PD is in logarithmic scale, and is the same as that plotted in Figure \ref{fig:PD_vs_mass}. For sources with widely varying inclination measurements (NGC 4151, 4U 1957+115) or only possible ranges (4U 1630-47) the full range of estimates is included in the error bars, with the data point placed in the center of the error bar. We use inclination estimates that were arrived independently of the IXPE polarization results. For LMC X-1 no inclination of the compact object is available, so we use the orbital inclination. For Cygnus X-1 the inclination range is from the orbital inclination estimate to the highest inclination estimate from spectral fitting. For Cygnus X-3 we use the jet inclination rather than the orbital inclination of the system, as it is more likely to trace the true inclination of the accretion flow. MCG 5-23-16 and IC 4329A have only upper limits on the inclination, so they are plotted as such. Note the large cluster of blue points including both BHXBs in the hard state and the unobscured AGN, which are all classified as being in the hard state. The AGN observed by IXPE all have accretion rates in the range $0.02-0.1\dot{M}_{Edd}$, which predicts they should be in the hard state \citep{esin1997,narayan2008}. The soft state BHXBs are clustered below the hard state sources in PD.  Cyg X-3 is clearly an outlier, having a PD similar to the other obscured sources but with much lower inclination.

There appears to be a correlation between the PD and inclination. We performed a linear least-squares regression to the PD and inclination data of our sources and found a p-value of 0.38, so this correlation is not statistically significant. If we exclude Cygnus X-3 from the regression, we find a p-value of 0.003. Grouping the PDs of the hard and soft state sources together and the PDs of the obscured sources together, we obtained a p- value of $8\times10^{-4}$ from a two sample KS test. This indicates that the distributions of obscured and unobscured source PDs differ to a statistically significant extent. In these tests we used the upper limits as the values for sources without detected polarization.

\begin{figure*}
\centering
\includegraphics[width=\textwidth]{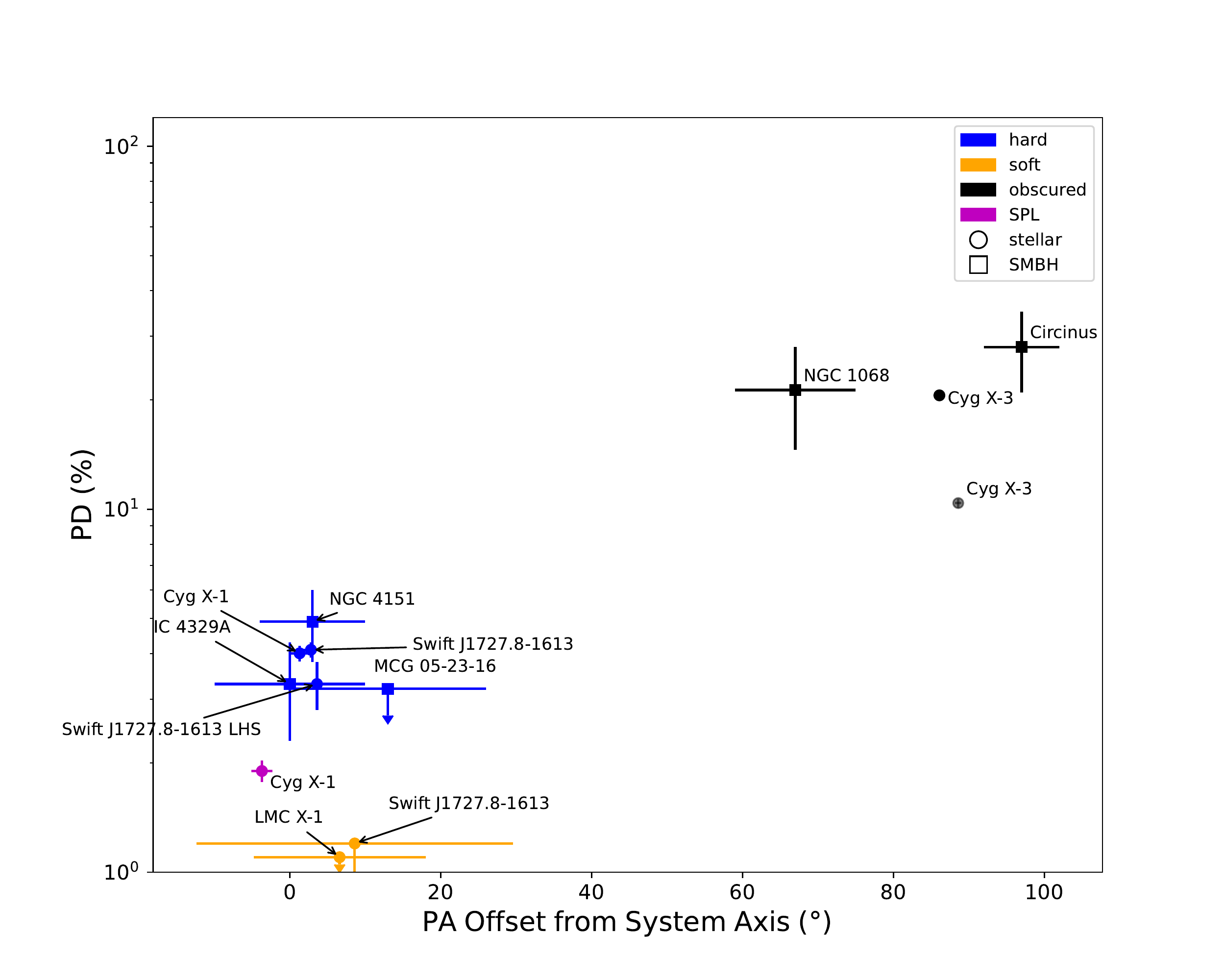}
\caption{Plot of the logarithm of the PD for each source vs the offset between the PA and the main axis of the system. Only sources with system axis position angle measurements are plotted. Blue denotes a hard state observation, magenta denotes an SPL observation, black denotes an obscured source and orange denotes a soft state observation. Polarization was not confidently detected for the soft state observations, thus the PA offset from system axis values should be regarded as tentative. Stellar mass black holes are plotted as circles, supermassive black holes are plotted as squares. Error bars on the system offset are derived from uncertainties on the PA only.}
\label{fig:PD_vs_offset}
\end{figure*}

For each source with system axis information, we subtracted the PA from the system axis to calculate the offset between the PA and the system axis. We then plot the PD vs the offset between the PA and the system axis in Figure \ref{fig:PD_vs_offset}. The uncertainty in the offset only includes the uncertainty in the PA, as for most sources the uncertainty in the system axis position angle was unavailable. Note that the sources appear to have a PA that is either perpendicular or parallel to the system axis; there are no intermediate cases. It is noticeable that the AGNs predicted to be in the hard state by their Eddington ratio have similar PA offsets from the system axis to the stellar black holes confirmed to be in the hard state. These results are tentative evidence that stellar and supermassive black holes have similar accretion geometry.

Doing a linear regression to the data in this plot, we find a p-value of $1.8\times10^{-4}$. Grouping the PA-system axis offsets of the obscured sources in one group and the unobscured sources in another group, a 2 sample KS test produces a p-value of 0.012. The difference in offset between the unobscured and obscured sources is thus statistically significant. In these calculations, we also use the upper limits as the PD values for sources without a polarization detection.

The soft state sources have PAs aligned with the system axis similarly to the hard state sources; however it must be noted that both soft state sources with system axis information have only upper limits on their PD, so their PA measurements must be treated with caution. The same follows for MCG 05-23-16, which appears to have PAs similar to the hard state sources, but also only has an upper limit on the PD. Overall these results are further evidence for shared accretion geometry in stellar and supermassive black holes, though this must be regarded as particularly tentative in the case of the sources with only upper limits on the PD.

\section{Discussion}\label{sec:discussion}
The most noteworthy feature of these data is that the AGN with SMBHs predicted to be in the hard state due to their low accretion rates share similar polarization properties with the hard state stellar black holes. They both have PDs of a few percent and PAs parallel to the system axis. They can be seen as the blue clump of points in the lower left of Figure \ref{fig:PD_vs_offset}. These are sources for which our view of the corona is unobstructed, and the similar polarization properties imply the coronal geometry is similar in both stellar and supermassive black holes in the hard state. This combined with the lack of any dependence of polarization properties on mass indicates that stellar and supermassive black holes share a common accretion geometry.

\begin{figure*}
\centering
\includegraphics[width=\linewidth]{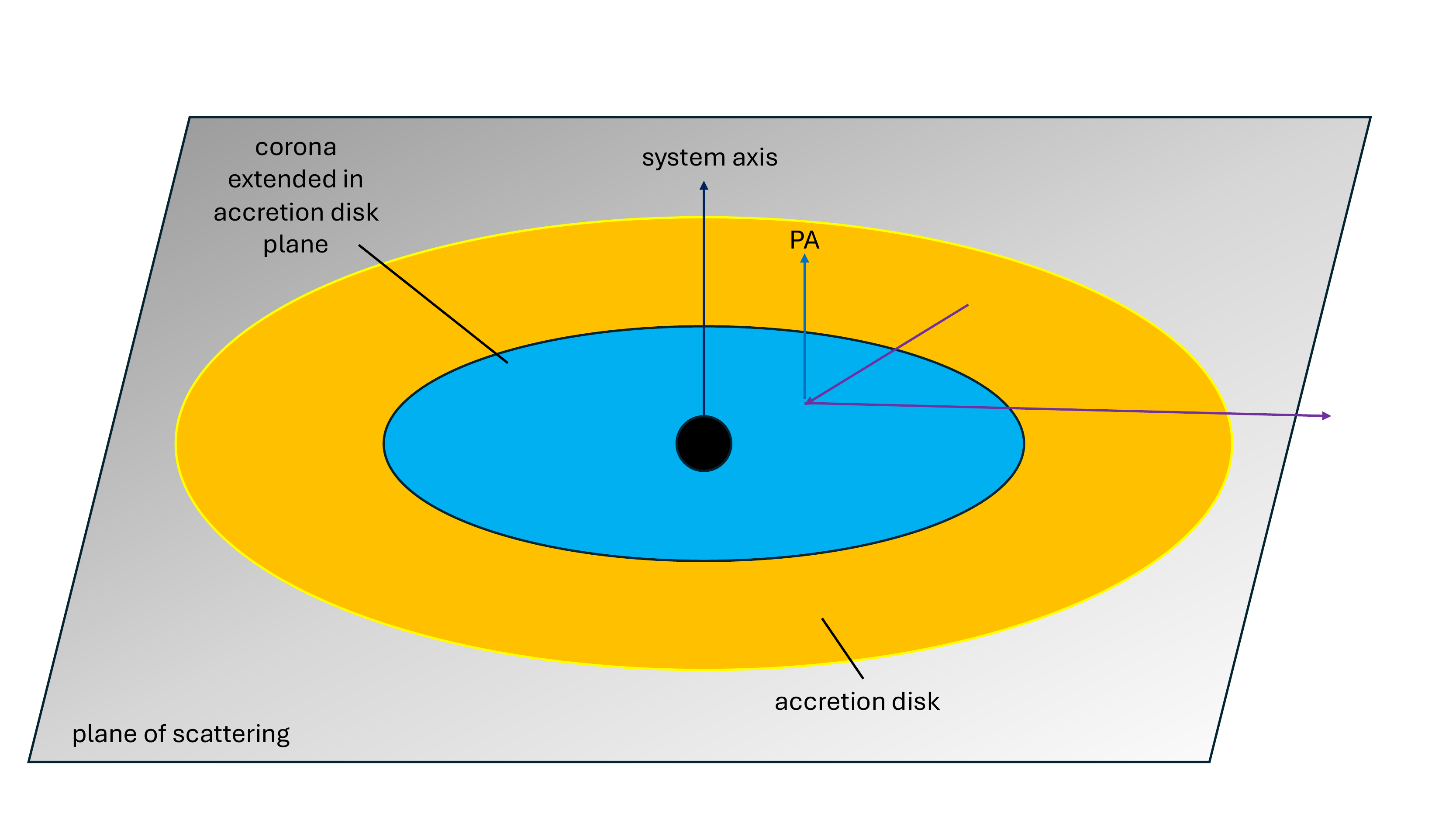}
\caption{Diagram of the polarization behavior of unobscured black holes in the hard state. Orange represents the accretion disk, blue represents the corona. The purple lines indicate the path of photons. Photons from the accretion disk inversely Compton scatter in the corona towards our line of sight. If the corona is extended along the plane of the accretion disk, then the polarization angle is oriented perpendicular to that plane, or parallel to axis of symmetry of the system. In this case the polarization vector is oriented vertically.}
\label{fig:hard_state}
\end{figure*}

Compton scattering of unpolarized light results in polarization perpendicular to the plane of scattering. In the hard state sources, we see directly down to the inner accretion flow and the compact object. In these inner reaches, the corona dominates over the accretion disk. The fact that the PA in the hard state sources is approximately aligned with the system axis indicates that the corona, where the scattering occurs, is extended in the direction perpendicular to the system axis, i.e. along the accretion disk  (see Figure \ref{fig:hard_state}). This geometry appears to hold both for stellar and supermassive accreting black holes in the the hard state.

For soft state sources, we also see directly down to the inner accretion flow and the compact object, but the accretion disk dominates over the corona. For the soft state sources with system axis information, the reported PAs appear to be aligned with the system axis, possibly indicating the scattering plane is extended along the accretion disk plane as well. We could associate this scattering plane with the accretion disk itself, but since we only have upper limits on the PD for the soft state sources with system axis information, we cannot come to any definite conclusions at this point.

\begin{figure*}
\centering
\includegraphics[width=\linewidth]{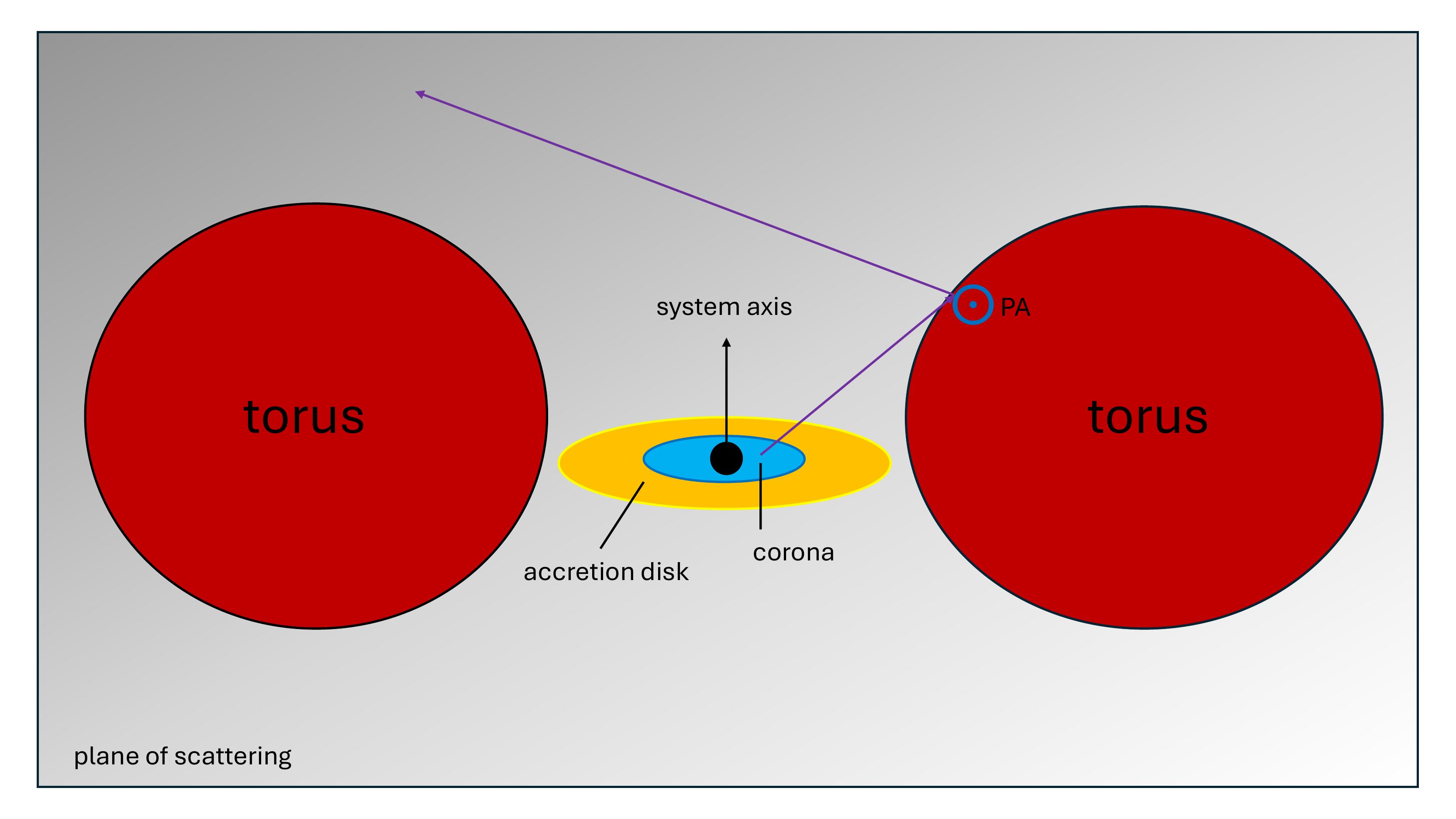}
\caption{Diagram of the polarization behavior of obscured black holes. Blue represents the corona, orange represents the accretion disk, red represents the obscuring torus. The purple lines indicate the path of photons. Photons from the corona Compton scatter off of the torus and into the line of sight. If the torus is extended in the vertical direction, then the polarization angle is oriented perpendicular to that plane, or perpendicular to axis of symmetry of the system. In this case, the polarization vector is pointed out of the page.}
\label{fig:obscured_state}
\end{figure*}

In the obscured sources, the PD is much higher than in the soft or hard state sources. We do not see the inner accretion flow directly in obscured sources, but rather X-rays scattered from material farther out in the accretion flow. Blockage of the central source implies that the obscuring material has a large optical depth. The photoelectric absorption cross-section is larger than the Compton cross-section in the IXPE band. This indicates that the X-ray photons likely are absorbed before they can complete more than a single Compton scattering in the obscuring material. Therefore, we likely observe photons that have only undergone a single Compton scattering. The fact that the PA in the obscured sources is approximately perpendicular to the system axis indicates that the scattering material is extended in the direction parallel to the system axis, i.e. perpendicular to the accretion disk (see Figure \ref{fig:obscured_state}). This geometry appears to hold both for stellar and supermassive accreting black holes where the central object is obscured. This could be caused by a funnel or torus of gas surrounding the central object, as is hypothesized to occur in AGN. To obscure our view of the central source, the opening angle of the funnel/torus must be less than the inclination of the line of sight to the source. The low inclination of Cygnus X-3 implies that the throat of the funnel surrounding it is quite narrow.

\section{Conclusion}\label{sec:conclusion}

We have shown that the polarization properties of stellar mass black holes and supermassive black holes are similar to each other, which constitutes tentative evidence that they share a common coronal geometry. Further observations of more black hole targets are needed to confirm whether this similarity continues to hold with additional data. Currently we do not possess any solid detections of polarization in soft state sources with known system axis orientations. We also lack polarization measurements for AGN that would be predicted to be in an equivalent of the soft state. Deeper IXPE observations of stellar black holes in the soft state with known system axis orientation as well as IXPE observations of AGN with  high accretion rates (and so presumably in the equivalent of a soft state) are therefore well-motivated.

The similar polarization properties of the stellar and supermassive black holes raise the possibility that we could use the high signal-to-noise IXPE observations afforded by Galactic black holes to inform our understanding of the accretion geometry of AGN, of which high SNR polarization measurements are much more difficult to obtain.

\section*{Acknowlegements}

We thank the anonymous referee for useful comments that have helped improve this paper.

I. L. was supported by the NASA Postdoctoral Program at the Marshall Space Flight Center, administered by Oak Ridge Associated Universities under contract with NASA.

The Imaging X-ray Polarimetry Explorer (IXPE) is a joint US and Italian mission. The US contribution is supported by the National Aeronautics and Space Administration (NASA) and led and managed by its Marshall Space Flight Center (MSFC), with industry partner Ball Aerospace (contract NNM15AA18C). The Italian contribution is supported by the Italian Space Agency (Agenzia Spaziale Italiana; ASI) through contract ASI-OHBI-2017-12-I.0, agreements ASI-INAF-2017-12-H0 and ASI-INFN-2017.13-H0, and its Space Science Data Center (SSDC), and by the Istituto Nazionale di Astrofisica (INAF) and the Istituto Nazionale di Fisica Nucleare (INFN) in Italy. This research used data products provided by the IXPE Team (MSFC, SSDC, INAF, and INFN) and distributed with additional software tools by the High-Energy Astrophysics Science Archive Research Center (HEASARC), at NASA Goddard Space Flight Center (GSFC).

\bibliography{bh_meta}{}
\bibliographystyle{aasjournal}

\end{document}